\documentclass[%
 aps, prl,
 reprint,
 twocolumn,
superscriptaddress,
 amsmath,amssymb, amsfonts,
]{revtex4-2}

\usepackage{graphicx}
\usepackage{dcolumn}
\usepackage{bm}
\usepackage{physics}
\usepackage{xcolor}
\definecolor{linkblue}{HTML}{2364AA}
\definecolor{linkgold}{HTML}{C1900B}
\definecolor{linkred}{HTML}{B52639}
\usepackage{hyperref}
\hypersetup{
	colorlinks = true ,
	linkcolor  = linkred ,
	urlcolor   = linkblue ,
	citecolor = red
} 
 
\newenvironment{psmallmatrix}
  {\left(\begin{smallmatrix}}
  {\end{smallmatrix}\right)}

\usepackage{tikz}
\usetikzlibrary{arrows}
\usepackage{floatrow}
\usepackage{graphicx}
\usepackage[label font=bf, position=top, caption=false]{subfig}

\usepackage{comment}

\begin{document}

\preprint{APS/123-QED}

\title{Interference-based universal decoupling and swapping for multimode bosonic systems}

\author{Mengzhen Zhang}
\affiliation{Departments of Applied Physics and Physics, Yale University, New Haven, CT 06520, USA}
\affiliation{Yale Quantum Institute, Yale University, New Haven, CT 06520, USA}
\affiliation{Pritzker School of Molecular Engineering, University of Chicago, Chicago, IL 60637, USA}
\author{Shoumik Chowdhury}
\affiliation{Departments of Applied Physics and Physics, Yale University, New Haven, CT 06520, USA}
\affiliation{Yale Quantum Institute, Yale University, New Haven, CT 06520, USA}
\author{Liang Jiang}
\affiliation{Departments of Applied Physics and Physics, Yale University, New Haven, CT 06520, USA}
\affiliation{Yale Quantum Institute, Yale University, New Haven, CT 06520, USA}
\affiliation{Pritzker School of Molecular Engineering, University of Chicago, Chicago, IL 60637, USA}

\date{October 17, 2020}%

\begin{abstract}
Beam-splitter operations are widely used to process information encoded in bosonic modes. In hybrid quantum systems, however, it might be challenging to implement a reliable beam-splitter operation between two distinct bosonic modes. Without beam-splitters, some basic operations such as decoupling modes and swapping states between modes can become highly non-trivial or not feasible at all.
In this work, we develop novel interference-based protocols for decoupling and swapping selected modes of a multimode bosonic system without requiring beam-splitters. Specifically, for a given generic coupler characterized by a Gaussian unitary process, we show how to decouple a single mode or swap any pair of modes with a constant depth sequence of operations, while maintaining the coupling for the remaining system. These protocols require only multiple uses of the given coupler, interleaved with single-mode Gaussian unitary operations, and thus enable efficient construction of operations crucial to quantum information science, such as high-fidelity quantum transduction. Our results are directly derived from fundamental physical properties of bosonic systems and are therefore broadly applicable to various existing platforms.
\end{abstract}

\maketitle


\textit{Introduction.}
Hybrid quantum networks are a promising platform for building large-scale modular quantum computers \cite{kimble2008quantum}, where bosonic quantum information processing plays a crucial role. A key requirement for hybrid bosonic quantum information processing is the ability to precisely engineer deterministic interactions between desired modes of the hybrid system.

For any given task, certain interactions are naturally more desirable than others. Within the context of quantum state transfer and quantum transduction, for instance, the \textit{swapping operation} is of paramount importance, allowing us to faithfully exchange quantum information between one bosonic mode and another. The development of robust quantum transducers is an active area of research, and there have recently been many notable advances in microwave-to-optical conversion \cite{hafezi2012atomic, bochmann2013nanomechanical, andrews2014bidirectional, tian2015optoelectromechanical, hisatomi2016bidirectional,rueda2016efficient, vainsencher2016bi, higginbotham2018harnessing}, microwave-to-mechanical/spin-wave conversion \cite{palomaki2013coherent, zhang2014strongly, tabuchi2015coherent}, and the transfer of quantum information between processor and memory modes for storage \cite{julsgaard2004experimental, sherson2006light, rabl2006hybrid, stannigel2010optomechanical, grezes2014multimode}. 

Oftentimes, desired quantum operations can be accompanied by unwanted interactions, such as those resulting the coupling between the system modes of interest and the environment. In these cases, we would ideally like to have a \textit{decoupling operation} to remove the effect of such couplings and thus protect the system from decoherence. In bosonic systems, this has been heavily investigated in the contexts of quantum error correction \cite{Ofek16, HuL19, Campagne19} and bosonic dynamical decoupling (DD) \cite{arenz2017dynamical, Heinze19}.


In consideration of the importance of \textit{decoupling} and \textit{swapping} operations, we focus here on the efficient construction of them. It is notable that although there have been a number of previous efforts in this direction as mentioned above, many of them face significant drawbacks. For example, DD is resource-demanding and requires a long sequence of controls, while giving only an approximate decoupling. More importantly, quantum operations coupling disparate bosonic modes in a hybrid system will inevitably suffer from the presence of sideband modes \cite{andrews2014bidirectional, rueda2016efficient, soltani2017efficient}. This leads to the general unavailability of beam-splitter interactions between arbitrary bosonic modes (e.g. microwave-to-optical, or microwave-to-mechanical). As a consequence, many known theorems and protocols from quantum optics for constructing \cite{dutta1995real, de2006symplectic} and decomposing \cite{Braunstein05,de2006symplectic} arbitrary Gaussian unitary operations --- which all require beam-splitters --- cannot be applied here to construct decoupling or swapping operations.

One recent attempt to overcome the limitations above is the adaptive protocol -- inspired by quantum teleportation -- of Ref. \cite{aqt2018}. Here, a given multimode bosonic interaction among all the modes of the system (including the undesired sideband modes) can be converted to an interaction involving only the desired modes, using only auxiliary single-mode Gaussian operations and classical communications. However, the ideal implementation of this protocol requires impractical resources such as infinite squeezing and perfect homodyne measurements.   


Therefore, one need a new scheme to address the aforementioned problems without requiring a large resource overhead. In this work, we develop a novel interference-based scheme for constructing universal decoupling and swapping operations. The structure of the scheme resembles that of the protocol demonstrated in Ref. \cite{2018hklau}; it requires only: (a) multiple copies of a given multimode interaction, and (b) free access to single-mode Gaussian unitary control operations (i.e. phase-shifting and/or single-mode finite squeezing). However, unlike Ref. \cite{2018hklau}, our scheme is derived  from fundamental physical properties of bosonic systems, and is thus applicable to arbitrary multi-mode situations. Specifically, we find that the coupling between an arbitrary pair of selected quadratures can be removed via interference. This is done by implementing the given bosonic interaction twice, interspersed with the local (i.e. single-mode) control operations. By constructing an inductive multi-pass sequence of the form above, we can then successively remove all unwanted coupling terms quadrature-by-quadrature. 

Based on these general ideas, we can either sequentially decouple an arbitrary number of modes, or swap two chosen modes from the multimode system, using a constant-depth sequence that is independent of the number of modes. We will start with examples in two-mode systems:

\textit{Two-mode decoupling.} Gaussian unitary physical processes involving linearly-coupled bosonic modes are completely determined by the change of the expectation values of the quadrature operators before and after the interaction. Specifically, we can organize such a transformation into a $2N\times 2N$ real symplectic matrix $\bm{S}$ mapping $\vu{x}\to\bm{S}\vu{x}$ where $N$ is the number of modes involved and $\vu{x} := (\hat{q}_1, \hat{p}_1, \ldots, \hat{q}_N, \hat{p}_N)^T$ is a collection of their respective quadrature operators \cite{weedbrook2012gaussian} (see Supplementary Material for definitions and conventions \cite{supplementary}). Without loss of generality, we can work entirely in terms of these symplectic matrices. 

Let us investigate a generic bosonic interaction process involving only two linearly-coupled modes. This can be described by a $4\times 4$ real symplectic matrix 
\begin{equation}
\bm{S} = \begin{pmatrix}
S_{11} & S_{12} & S_{13} & S_{14}\\
S_{21} & S_{22}  & S_{23} & S_{24}\\
S_{31} & S_{32} & S_{33} & S_{34}\\
S_{41} & S_{42} & S_{43} & S_{44}\\
\end{pmatrix}.
\end{equation} 

\begin{figure}[t]
\centering
\subfloat[\textbf{Sequence for two-mode decoupling}\hfill\textcolor{white}{.}]{\includegraphics[width=\textwidth]{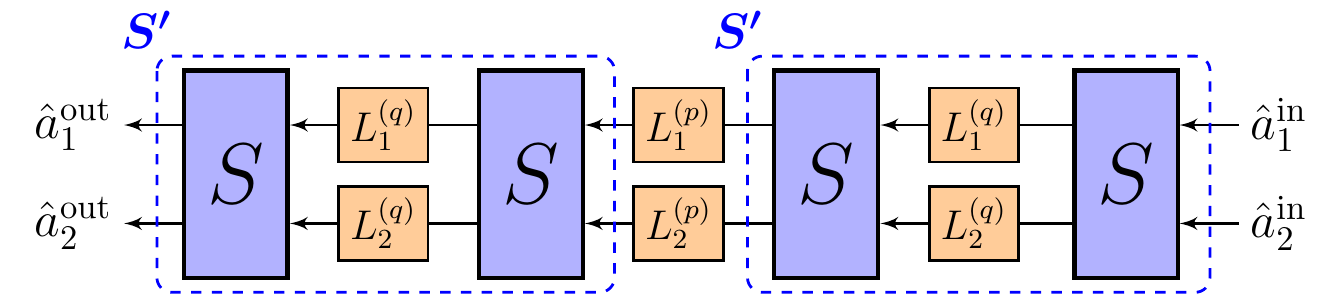}}

\medskip
\subfloat[\textbf{Sequence for two-mode perfect transduction}\hfill\textcolor{white}{.}]{\includegraphics[width=\textwidth]{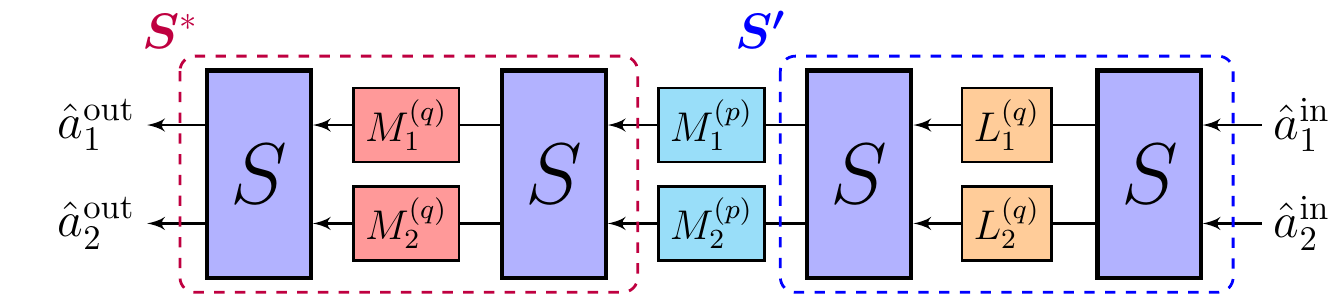}}
\caption{(a) To decouple two interacting modes, we construct a sequence involving four copies of the interaction $\bm{S}$, interspersed by local operations. The operations $\bm{L}_k^{(q)}$ are sandwiched between two copies of $\bm{S}$ to construct an effective interaction $\bm{S'}$ with the $\hat{q}_1$-quadrature decoupled from mode 2. We then repeat this step recursively using $\bm{S'}$ and modified local operations $\bm{L}_k^{(p)}$ to yield a net process with both $\hat{q}_1$ and $\hat{p}_1$ quadratures decoupled. (b) To construct a swap operation, we instead use two different interactions $\bm{S}^\ast$ and $\bm{S'}$, where $\bm{S}^\ast$ is constructed so that $\hat{q}_2^{\rm out} = \hat{p}_1^{\rm in}$ and $\hat{q}_1^{\rm in}$ is only present in $\hat{p}_2^{\rm out}$. Using interference, we can then cancel the contamination of $\hat{p}_2^{\rm out}$ by the other input quadratures besides $\hat{q}_1^{\rm in}$ to yield (up to local operations) a perfect two-mode swap.}
\label{fig:two-mode}
\end{figure}

The protocol that we develop is based on running this interaction multiple times to form a multi-pass sequence as shown in Fig. \ref{fig:two-mode}(a). After each pass, we can apply local (i.e. single-mode) operations to tune the interference between quadratures.  It turns out that by carefully choosing these local operations, we can end up with a net process where the two modes are decoupled from one another. To see this, let us choose:

\begin{equation}
\big(\bm{L}_k^{(q)}\big)_{ij} = \frac{(-1)^{i+1} S_{1 \underline{i}} S_{\bar{j} 1}}{ (S_{\bar{i} 1})^2 + (S_{\underline{i}1})^2}  + 
\frac{(-1)^{j+1}  S_{1 \bar{i}} S_{\underline{j} 1}}{ (S_{1\bar{j}})^2 + (S_{1\underline{j}})^2},
\label{eq:2modelocal-1}
\end{equation}
with $i, j, k \in \{1, 2\}$ and where $\bar{i} = 2k-1$, $\underline{i}=2k$ if $i = 1$, and 
$\bar{i} = 2k $, $\underline{i}=2k-1$ if $i = 2$. Then, we can check by direct calculation that  
\begin{equation}
\bm{S'}=\bm{S} \begin{pmatrix}
\bm{L}_1^{(q)} & 0 \\
0 & \bm{L}_2^{(q)}  \\
\end{pmatrix} \bm{S} = \begin{pmatrix}
0 & 1 & 0 & 0\\
-1 & S'_{22}  & S'_{23} & S'_{24}\\
0 & S'_{32} & S'_{33} & S'_{34}\\
0 & S'_{42} & S'_{43} & S'_{44}\\
\end{pmatrix},
\end{equation}
is of the form above, which corresponds to an effective interaction with no coupling between the $\hat{q}_1$ quadrature and quadratures $\hat{q}_2, \hat{p}_2$ for mode 2. (Note that the construction of the local operations throughout the text is not unique, and one can even modify them to relax the requirement of squeezing. See Supplementary Material \cite{supplementary} for more details). It now remains only to decouple the $\hat{p}_1$ quadrature, which is achieved via
\begin{equation}
\big(\bm{L}_2^{(p)}\big)_{ij} = \frac{(-1)^{i+1} S_{2 \underline{i}}' S_{\bar{j} 2}'}{ (S_{\bar{i} 1}')^2 + (S_{\underline{i}1}')^2}  + 
\frac{(-1)^{j+1}  S_{2 \bar{i}}' S_{\underline{j} 2}'}{ (S_{1\bar{j}}')^2 + (S_{1\underline{j}}')^2},
\label{eq:2modelocal-2}
\end{equation} 
and $\bm{L}^{(p)}_1 = -\bm{\omega} = \begin{psmallmatrix}
0 & -1 \\ 1 & 0
\end{psmallmatrix}$, with $i,j\in \{1,2\}$  \footnote{We  note that $\bm{\omega} = \bm{R}(\pi/2)$ and is thus a passive local operation.}. If we then explicitly calculate
\begin{equation}
\bm{S}^{\rm dc} =\bm{S'} \begin{pmatrix}
\bm{L}_1^{(p)} & 0 \\
0 & \bm{L}_2^{(p)}  \\
\end{pmatrix} \bm{S'} = \begin{pmatrix}
0 & 1 & 0 & 0\\
-1 & S^{\rm dc}_{22}  & 0 & 0\\
0 & 0 & S^{\rm dc}_{33} & S^{\rm dc}_{34}\\
0 & 0 & S^{\rm dc}_{43} & S^{\rm dc}_{44}\\
\end{pmatrix},
\end{equation}
we see that $\bm{S}^{\rm dc}$ has a block diagonal structure, reflecting the lack of coupling between the two modes in the net process. Note that, for now, we have assumed that the original matrix $\bm{S}$ is \textit{generic}, i.e. meaning that the denominator of Eq. \eqref{eq:2modelocal-1} is always non-vanishing. The non-generic cases are treated in Supplementary Material \cite{supplementary}.

\textit{Two-mode quantum transduction.} 
With minor changes to the intermediate local operations, our protocol can be directly applied to constuct perfect quantum transducers. Let us define the local operations $\bm{M}_k^{(q)}$ via:

\begin{equation}
\big(\bm{M}_k^{(q)}\big)_{ij} =
\frac{(-1)^{i+1} S_{3 \underline{i}} S_{\bar{j} 1}}{ (S_{\bar{i} 1})^2 + (S_{\underline{i}1})^2} -  \frac{(-1)^{j+1}  S_{3 \bar{i}} S_{\underline{j} 1}}{ (S_{3\bar{j}})^2 + (S_{3\underline{j}})^2},
\end{equation}
which are chosen to ensure that $\hat{q}_2^{\rm out} = \hat{p}_1^{\rm in}$. Once again, $i, j, k \in \{1, 2\}$ and we have $\bar{i} = 2k-1$, $\underline{i}=2k$ if $i = 1$, and 
$\bar{i} = 2k $, $\underline{i}=2k-1$ if $i = 2$. Using these operations, we can then construct
\begin{equation}
\bm{S}^\ast=\bm{S} \begin{pmatrix}
\bm{M}_1^{(q)} & 0 \\
0 & \bm{M}_2^{(q)}  \\
\end{pmatrix} \bm{S} = \begin{pmatrix}
0 & S_{12}^\ast & S_{13}^\ast & S_{14}^\ast\\
0 & S_{22}^\ast  & S_{23}^\ast & S_{24}^\ast\\
0 & 1 & 0 & 0\\
-1 & S_{42}^\ast & S_{43}^\ast & S_{44}^\ast\\
\end{pmatrix},
\end{equation}
which indeed has the desired form. Observe from the structure of this matrix that $\hat{q}_1^{\rm in}$ is fully transferred to $\hat{p}_2^{\rm out}$ though additional contamination from the other quadratures is still present. Our goal is then to cancel these contributions by sandwiching local operations between $\bm{S}^\ast$ and $\bm{S'}$, as shown in Fig.~\ref{fig:two-mode}(b). Here, we choose the local operations
\begin{equation}
\big(\bm{M}_2^{(p)}\big)_{ij} =
\frac{(-1)^{i+1} S^\ast_{4 \underline{i}} S_{\bar{j} 2}'}{ (S_{\bar{i} 2}')^2 + (S_{\underline{i}2}')^2} -  \frac{(-1)^{j+1}  S^\ast_{4 \bar{i}} S_{\underline{j} 2}'}{ (S_{4\bar{j}}^{\ast})^2 + (S_{4\underline{j}}^{\ast})^2}.
\end{equation}
and $\bm{M}_1^{(p)} = -\bm{\omega} = \begin{psmallmatrix}
0 & -1 \\ 1 & 0
\end{psmallmatrix}$. 
By explicit calculation, the matrix $\bm{S}^{\rm td}$ describing the net process will be of the form
\begin{equation}
\bm{S}^{\rm td} =\bm{S}^\ast \begin{pmatrix}
\bm{M}_1^{(p)} & 0 \\
0 & \bm{M}_2^{(p)}  \\
\end{pmatrix} \bm{S'} = \begin{pmatrix}
0 & 0 & S^{\rm td}_{13} & S^{\rm td}_{14}\\
0 & 0  & S^{\rm td}_{23} & S^{\rm td}_{24}\\
0 & 1 & 0 & 0\\
-1 & S^{\rm td}_{42} & 0 & 0
\end{pmatrix},
\label{eq:2-mode-std}
\end{equation}
which, up to local operations, is equivalent to swapping the two modes (referred to as \textit{perfect transduction}).

\textit{Multimode decoupling.} 
It turns out that the decoupling and transduction protocols described above are not accidental, but rather stem from fundamental physical properties of bosonic systems -- they thus readily generalize to the multimode case. To see why, it will be helpful to first introduce a geometric interpretation of the symplectic matrices, whose rows (or columns) form an orthonormal symplectic basis \cite{de2006symplectic}. Specifically, for an arbitrary  $2N\times 2N$ symplectic matrix $\bm{S}$, we can denote its rows via $\bm{S} = (\vb{u}_1, \vb{v}_1, \ldots, \vb{u}_N, \vb{v}_N)^T$ and its columns via $\bm{S} = (\vb{x}_1, \vb{y}_1, \ldots, \vb{x}_N, \vb{y}_N)$, 
where $\vb{u}_i, \vb{v}_i, \vb{x}_i$ and $\vb{y}_i$  are $2N$-dimensional vectors, with $1\le i \le N$. 

Since any unitary physical process described by a symplectic matrix $\bm{S}$ must preserve the canonical commutation relations, the matrix $\bm{S}$ must satisfy the defining condition $\bm{S}\bm{\Omega} \bm{S}^T = \bm{\Omega}$, where $\bm{\Omega} = \bigoplus_{i=1}^N \bm{\omega} = \mbox{diag}(\bm{\omega}, \dots, \bm{\omega})$, and where $\bm{\omega} = \begin{psmallmatrix}
0 & 1 \\ -1 & 0
\end{psmallmatrix}$ is called the \textit{symplectic form}. Therefore, the matrix $\bm{\Omega}$ can be considered a `local' operation as defined: it simply corresponds to a $\pi/2$ phase-shift on each mode. This condition gives us an explicit set of orthogonality relations between the rows and columns: $\vb{u}_i^T \vb{\Omega} \vb{u}_j=\vb{v}_i^T  \vb{\Omega} \vb{v}_j=\vb{x}_i^T \vb{\Omega} \vb{x}_j=\vb{y}_i^T \vb{\Omega} \vb{y}_j=0$, and $\vb{u}_i^T \vb{\Omega} \vb{v}_j=\vb{x}_i^T \vb{\Omega} \vb{y}_j = \delta_{ij}$,  where $i,j \in \{1,2, \dots, N\}$. Comparing these relations to the similar properties of orthogonal matrices, one can thus think of symplectic matrices as geometric transformations on the spaces spanned the row (or column) vectors. 

\begin{figure}[t]
\centering
\subfloat[\textbf{``Sandwich operation''}\hfill\textcolor{white}{.}]{\includegraphics[width=0.58\linewidth]{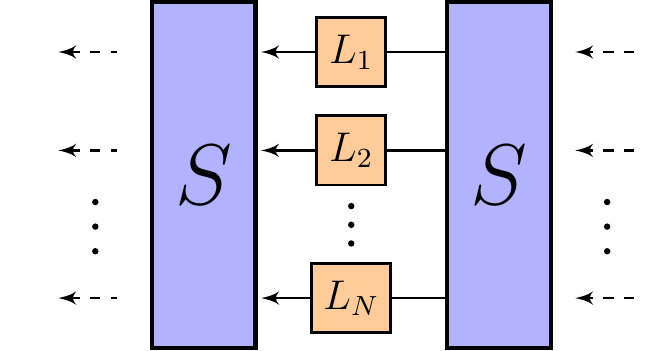}}\hfill\subfloat[\textbf{Decoupled mode}]{\includegraphics[width=0.41\linewidth]{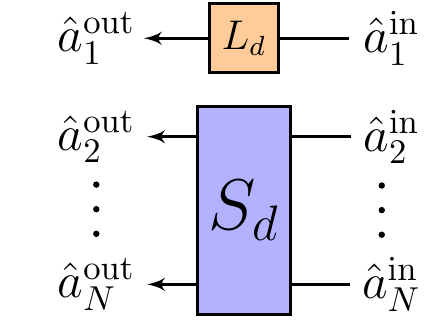}}

\medskip
\subfloat[\textbf{Generic sequence for decoupling a single mode}\hfill\textcolor{white}{.}]{\includegraphics[width=\linewidth]{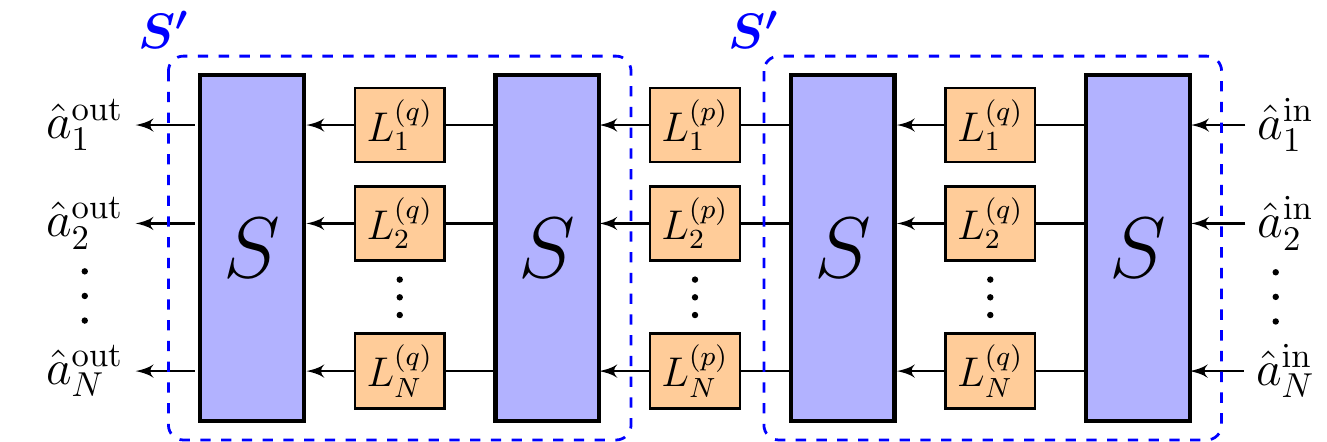}}
\caption{(a) We define a ``sandwich operation'' to denote two copies of the interaction $\bm{S}$ interspersed with local operations $\bm{L}_1, \bm{L}_2, \ldots, \bm{L}_N$. (b) A decoupled mode has no effective interaction with the remaining modes of the system; here we show $\hat{a}_1^{\rm in}$ decoupled so that it is equivalent to $\hat{a}_1^{\rm out}$ up to a local operation. (c) Our protocol demonstrates that a sequence of two sandwich operations interspersed by carefully chosen local operations can yield a net interaction with one mode decoupled. This process can then be repeated inductively to remove any number of coupling terms to unwanted modes.}
\label{fig:sketch}
\end{figure}

Let us consider the general decoupling protocol for multiple modes. As shown in Fig. \ref{fig:sketch}, we can decouple an individual mode in two steps (i.e. first decoupling the $\hat{q} -$quadrature and then the $\hat{p} -$quadrature). The motivation of each step is to build up a certain destructive interference between the quadratures using the geometric relations above. For concreteness, we will demonstrate how to decouple the first mode $\hat{a}_1$ from the others. Nevertheless, with different choice of local operations, the same steps can be used to decouple any mode. 

In the first recursive step, our goal is to construct an interaction $\bm{S'}=\bm{S} \bm{L}^{(q)}\bm{S}$ where the first  quadrature of the first mode is decoupled from the other modes (i.e. this means $\bm{S'}_{1j} = \bm{S'}_{j1} = 0 $ for $j > 2$). Explicitly, we have
\begin{equation}
    \bm{S'} = \begin{pmatrix}
    \text{---} & \vb{u}_1^T & \text{---} \\
    \text{---} & \vb{v}_1^T & \text{---} \vspace{-0.2cm} \\
    \vdots & & \vdots \\
    \text{---} & \vb{u}_N^T & \text{---} \\
    \text{---} & \vb{v}_N^T & \text{---}
\end{pmatrix}\bm{L}^{(q)}\begin{pmatrix}
    \vert & \vert & \cdots & \vert & \vert \\
    \vb{x}_1  & \vb{y}_1 & \cdots & \vb{x}_N & \vb{y}_N   \\
    \vert & \vert & \cdots & \vert & \vert
\end{pmatrix}
\end{equation}
using the row/column notation introduced before, where $\bm{L}^{(q)} = \text{diag}(\bm{L}_1^{(q)}, \ldots, \bm{L}_N^{(q)})$ is a series of local operations. We claim that if there exists an $\bm{L}^{(q)}$ such that $\bm{L}^{(q)}\vb{x}_1 =-\bm{\Omega} \vb{u}_1$, then this particular ``sandwich operation'' will decouple the $\hat{q}_1$ quadrature as desired. The reason for this is as follows: by the properties above, $\vb{x}_1$ is naturally orthogonal to each of the other columns except $\vb{y}_1$, and thus $\bm{L}^{(q)} \vb{x}_1$ will be orthogonal to each of the modified columns except for $\bm{L}^{(q)} \vb{y}_1$. Furthermore, since $\bm{L}^{(q)} \vb{x}_1 = -\bm{\Omega}\vb{u}_1$ by assumption, it will \textit{also} be orthogonal to each of the rows except for $\vb{v}_1$. Thus $\bm{S}\bm{L}^{(q)}\bm{S}$ will be of the expected form:
\begin{equation}
    \bm{S'} = \bm{S}\bm{L}^{(q)}\bm{S} = \begin{pmatrix}
      0 & 1 & 0 & \ldots & 0 \\
    -1 & S_{22}' & S_{23}' & \ldots & S_{2, 2N}' \\
    0 & S_{32}' & S_{33}'  & \ldots & S_{3, 2N}' \\
    \vdots & \vdots  & \vdots & \ddots & \vdots \\
    0 & S_{2N, 2}' & S_{2N, 3}'  & \ldots & S_{2N, 2N}' \\
    \end{pmatrix}
\end{equation}

With the $\hat{q}_1$-quadrature decoupled from the remaining modes, we now proceed to the second recursive step of our protocol to decouple $\hat{p}_1$. This involves repeating the technique above using $\bm{S'}$ instead, with some minor modifications to the local operations. Once again, let us denote the rows of this matrix by $\bm{S'} = (\bm{\alpha}_1, \bm{\beta}_1, \ldots, \bm{\alpha}_N, \bm{\beta}_N)^T$ and the columns by $\bm{S'} = (\bm{\chi}_1, \bm{\gamma}_1, \ldots, \bm{\chi}_N, \bm{\gamma}_N)$.
We want to build up a ``sandwich operation'' of the form $\bm{S}^{\rm dc} = \bm{S'}\bm{L}^{(p)}\bm{S'}$ such that the local operation $\bm{L}^{(p)}$ transforms the second column vector $\bm{\gamma}_1 \to \bm{L}^{(p)}\bm{\gamma}_1 = -2 (S'_{22}) \bm{\Omega} \bm{\alpha}_1 +\bm{\Omega} \bm{\beta}_1$ while also transforming the first column vector $\bm{\chi}_1 \to  \bm{L}^{(p)}\bm{\chi}_1 = -\bm{\Omega}\bm{\alpha}_1$ \footnote{This is possible if we let $\bm{L}^{(p)}_1 =  -\bm{\omega}$. This is purely a phase rotation, i.e. no squeezing necessary.}.

Since $\bm{L}^{(p)}\bm{\chi}_1$ and $\bm{L}^{(p)}\bm{\gamma}_1$ are linearly independent, the two-dimensional plane spanned by the pair of vectors $\bm{\Omega}\bm{\alpha}_1, \bm{\Omega}\bm{\beta}_1$ is identical to that spanned by the vectors $\bm{L}^{(p)}\bm{\chi}_1, \bm{L}^{(p)}\bm{\gamma}_1$. Consequently,  this plane is orthogonal to every other vector
$\bm{\Omega}\bm{\alpha}_j, \bm{\Omega}\bm{\beta}_j$, $\bm{L}^{(p)}\bm{\chi}_j,\bm{L}^{(p)}\bm{\gamma}_j$ for $j\geq 2$, as guaranteed by the geometrical relations above. Thus, the ``sandwich operation'' $\bm{S'}\bm{L}^{(p)}\bm{S'}$ will be:
\begin{equation}
    \bm{S}^{\rm dc} = \bm{S'}\bm{L}^{(p)}\bm{S'} = \begin{pmatrix}
      0 & 1 & 0  & \ldots & 0 \\
    -1 & S_{22}^{\rm dc} & 0 & \ldots & 0 \\
    0 & 0 & S_{33}^{\rm dc}  & \ldots & S_{3, 2N}^{\rm dc} \\
    \vdots & \vdots  & \vdots & \ddots & \vdots \\
    0 & 0 & S_{2N, 3}^{\rm dc}  & \ldots & S_{2N, 2N}^{\rm dc} \\
    \end{pmatrix}.
    \label{eq:multimode-sdc}
\end{equation}

Observe that $\bm{S}^{\rm dc}$ is exactly of the form shown in Fig. \ref{fig:sketch}(b): the first mode is decoupled from all the others. At this point, we can proceed inductively and apply the same protocol to the $N-1$ mode subblock of $\bm{S}^{\rm dc}$. Doing so, we can in principle decouple any number of modes from the system. 

\begin{figure}[t]
    \centering
    \includegraphics[width=\linewidth]{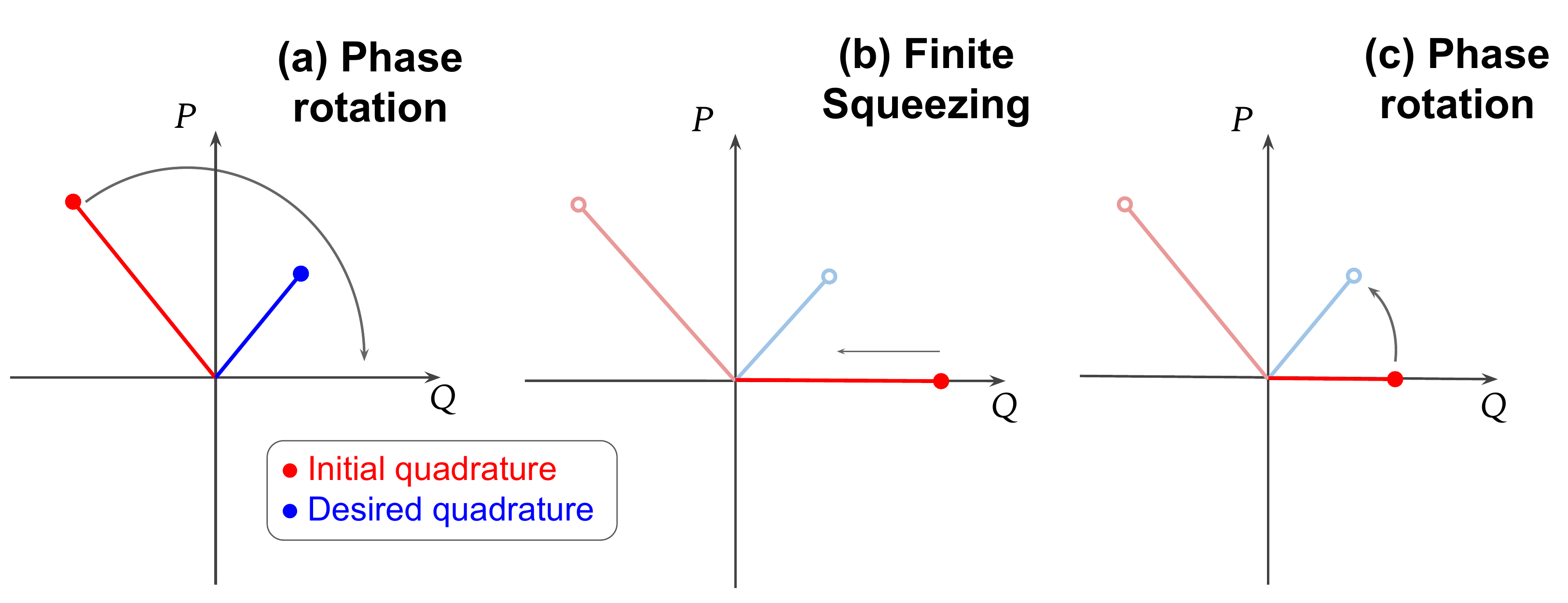}
    \caption{Geometric argument for the existence of local operations. Given any nonzero single-mode quadrature vector, it is possible to transform to another nonzero quadrature vector using only phase-space rotations and finite squeezing. }
    \label{fig:existence}
\end{figure}

It remains to be shown that the appropriate local operations $\bm{L}^{(q)}$ and $\bm{L}^{(p)}$ can always be constructed. By definition, each of them  is a direct sum of single-mode operations: e.g. $\bm{L}^{(q)} = \text{diag}(\bm{L}_1^{(q)}, \ldots, \bm{L}_N^{(q)})$, where $\bm{L}_i^{(q)}$ are $2\times 2$ matrices. Therefore, it suffices to show that we can transform any generic two-dimensional vector to another using just a single local operation. As demonstrated in Fig. \ref{fig:existence}, this is always generically satisfied: any required local operation can be realized using a sequence of three elementary local operations: (a) rotation to the $q-$axis, (b) dilation, and (c) rotation to the final direction. In the language of linear optics, rotation and dilation correspond to phase-shifting and finite squeezing, respectively. The existence of $\bm{L}^{(q)}$ and $\bm{L}^{(p)}$ is thus always guaranteed \textit{unless} either the initial or the final vector is the zero vector. We refer to \textit{generic} scattering matrices as those for which the local operations can be constructed (i.e. they contain no zero subblock in any quadrature vector). The procedure for handling the non-generic cases is discussed in Supplementary Material \cite{supplementary}. 

\begin{figure}[t]
\centering
\includegraphics[width=\textwidth]{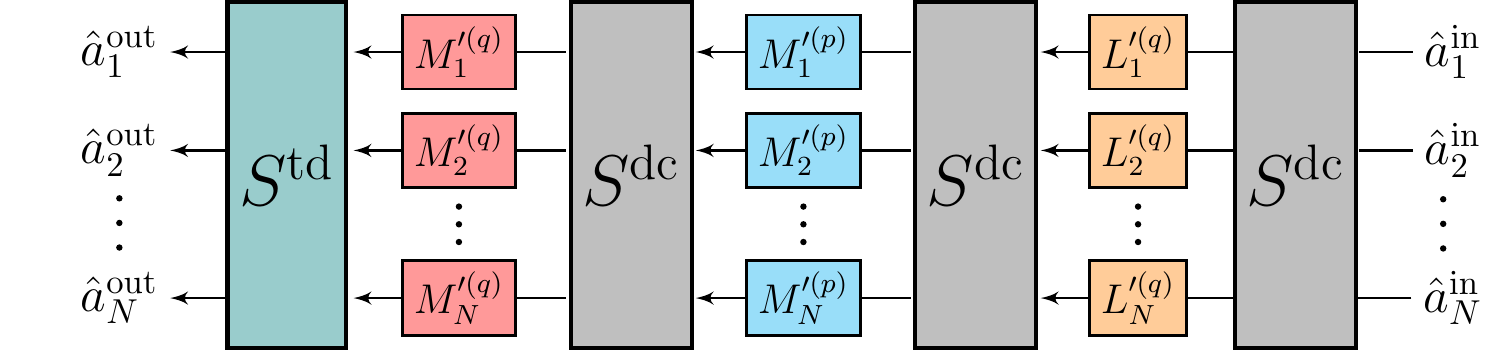}
\caption{In the multimode case, as in the two-mode case, we can apply our protocols to construct `transducer-type' interactions $\bm{S}^{\rm td}$ and `decoupling-type' interactions $\bm{S}^{\rm dc}$. By concatenating a transducer-type interaction with three decoupling-type interactions, as shown, we can realize an effective two-mode swap \textit{within} the multimode interaction for \textit{any} two given modes.}
\label{fig:transduction}
\end{figure}

\textit{Multimode quantum transduction.} With minor modifications, we can also generalize the quantum transduction protocol to multimode systems. Similar to the two-mode case, we can use four copies of a generic $\bm{S}$ to construct an effective `transducer-type' interaction of the form: \begin{equation}
\bm{S}^{\rm td}  = \begin{pmatrix}
0 & 0 & S^{\rm td}_{1,3} &\dots & S^{\rm td}_{1,2N}\\
\vdots & \vdots & \vdots &\ddots & \vdots\\
0 & 0  & S^{\rm td}_{2(N-1),3} &\dots & S^{\rm td}_{2(N-1),2N}\\
0 & 1 & 0 & \dots &0\\
-1 & S^{\rm td}_{2N,2} & 0 & \dots &0\\
\end{pmatrix}.
\label{eq:multimode-std}
\end{equation}
Here, we have constructed an `asymmetric swap' that transfers information from $\hat{a}_1^{\rm in}$ to $\hat{a}_N^{\rm out}$ but not vice versa. By combining this `transducer'-type interaction with the `decoupling'-type interaction in Eq. \eqref{eq:multimode-sdc}, however, it is possible to convert the asymmetric swap into a symmetric one. As shown in Fig.~\ref{fig:transduction}, we can combine  $\bm{S}^{\rm td}$ with three copies of $\bm{S}^{\rm dc}$ in order to successively remove coupling terms in the upper-right subblock of $\bm{S}^{\rm td}$. Using the decoupling result, we can find three local operations $\bm{L'}^{(q)}$, $\bm{M'}^{(p)}$, and $\bm{M'}^{(q)}$ such that the net process $\bm{\tilde{S}}$ has the form
\begin{align}
\begin{split}
     &\bm{\tilde{S}} = \big(\bm{S}^{\rm td} \bm{M'}^{(q)} \bm{S}^{\rm dc}\big) \bm{M'}^{(p)} \big(\bm{S}^{\rm dc} \bm{L'}^{(q)} \bm{S}^{\rm dc}\big)  \\
&= \begin{pmatrix}
0 & 0 & 0 & \dots & 0 &\tilde{S}_{1,2N-1} & \tilde{S}_{1,2N}\\
0 & 0 & 0 & \dots & 0 &\tilde{S}_{2,2N-1} & \tilde{S}_{2,2N}\\
0 & 0 & \ast & \dots & \ast &0 & 0\\
\vdots & \vdots & \vdots &\ddots &\vdots & \vdots & \vdots\\
0 & 0  & \ast &\dots &\ast & 0 & 0\\
\tilde{S}_{2N-1,1} & \tilde{S}_{2N-1,2} & 0 & \dots &0 & 0 & 0\\
\tilde{S}_{2N,1} & \tilde{S}_{2N,2} & 0 & \dots &0 & 0 & 0\\
\end{pmatrix},
\end{split}
\end{align}

\noindent which corresponds to a perfect swap (up to local operations) between the first and the last modes, with no coupling to the remaining $N-2$ modes. We can even repeat this process recursively to obtain swap operations between any arbitrary pairs of modes involved in the original interaction process $\bm{S}$. We note that $\bm{S}^{\rm dc}$ and $\bm{S}^{\rm td}$ can each be generated using four copies of $\bm{S}$, provided this interaction is generic (i.e. we can construct the local operations as needed). Thus, generically, we require a fixed overhead of just \textit{sixteen} copies of $\bm{S}$ to realize this generalized two-mode swap. More details are given in Supplementary Material \cite{supplementary}.

\textit{Discussion.} We note that the structure of our interference-based protocol closely resembles that of Ref. \cite{2018hklau}. In general, both schemes require local control over all involved modes. A key difference, however, is that our scheme works for $N > 2$ modes and can be used for multimode transduction \textit{and} decoupling \cite{andrews2014bidirectional, HanX20, Mirhosseini20, rueda2016efficient, soltani2017efficient}. Furthermore, we can realize a two-mode swap using at most only four copies of a generic transducer, whereas their scheme requires at most six. In the two-mode case, our scheme --- like that of Ref. \cite{2018hklau} --- can be adapted to require squeezing on \textit{only one} of the two modes. More generally, for the $N$-mode case, we can relax the squeezing requirement in each recursive step so that overall only $N-1$ modes require finite squeezing (see Supplementary Material \cite{supplementary}). 

While the qualification above holds for all generic scattering matrices, we have observed in numerical experiments that for certain practical systems (e.g. the optical-mechanical-microwave transducer of Ref. \cite{andrews2014bidirectional, HanX20, Mirhosseini20}), our protocols can be implemented without requiring \textit{any} local squeezing at all. There are also other situations where both quadratures of a selected mode can be decoupled simultaneously in a single sandwich operation, effectively halving the number of recursive steps required. The conditions for squeezing-free operation or simultaneous quadrature decoupling are the subject of ongoing investigation.


\textit{Conclusion.} In this work, we demonstrate novel interference-based protocols to decouple unwanted modes and efficiently construct high-fidelity swapping operations in a multimode bosonic system, assuming practical quantum resources.  These protocols are predicated on the fundamental physical properties of bosonic systems, and are thus universally applicable to arbitrary bosonic platforms. As a result, our widely available protocols may have broad applications in quantum information science.

\textit{Acknowledgement.} We thank Aashish Clerk, Hoi-Kwan Lau and Oskar Painter for stimulating discussions. We gratefully acknowledge support from the ARL-CDQI (W911NF-15-2-0067), ARO (W911NF-18-1-0020, W911NF-18-1-0212), ARO MURI (W911NF-16-1-0349), AFOSR MURI (FA9550-15-1-0015, FA9550-19-1-0399), DOE (DE-SC0019406), NSF (EFMA-1640959, OMA-1936118), and the Packard Foundation (2013-39273).

\bibliography{bibliography}

\end{document}


\preprint{APS/123-QED}

\title{Interference-based universal decoupling and swapping for multimode bosonic systems}

\author{Mengzhen Zhang}
\affiliation{Departments of Applied Physics and Physics, Yale University, New Haven, CT 06520, USA}
\affiliation{Yale Quantum Institute, Yale University, New Haven, CT 06520, USA}
\affiliation{Pritzker School of Molecular Engineering, University of Chicago, Chicago, IL 60637, USA}
\author{Shoumik Chowdhury}
\affiliation{Departments of Applied Physics and Physics, Yale University, New Haven, CT 06520, USA}
\affiliation{Yale Quantum Institute, Yale University, New Haven, CT 06520, USA}
\author{Liang Jiang}
\affiliation{Departments of Applied Physics and Physics, Yale University, New Haven, CT 06520, USA}
\affiliation{Yale Quantum Institute, Yale University, New Haven, CT 06520, USA}
\affiliation{Pritzker School of Molecular Engineering, University of Chicago, Chicago, IL 60637, USA}

\date{July 5, 2020}%

\maketitle

\section{Definitions}
The conventions used in this work closely follow the standard definitions for continuous-variable quantum information \cite{weedbrook2012gaussian}. For completeness, we will review the salient details below. 

We consider multimode systems comprised of $N$ bosonic modes, which correspond to $N$ pairs of bosonic field operators $(\hat{a}_1, \hat{a}_1^\dagger, \ldots, \hat{a}_N, \hat{a}_N^\dagger)^T \equiv \vu{a}$, where $[\hat{a}_j, \hat{a}_k^\dagger] = \delta_{jk}$. We can equivalently describe the system using quadrature operators $\hat{q}_k \equiv (\hat{a}_k + \hat{a}_k^\dagger)/\sqrt{2}$ and $\hat{p}_k \equiv i(\hat{a}_k^\dagger - \hat{a}_k)/\sqrt{2}$, which satisfy the canonical commutation relations. We also define the quadrature vector $\vu{x} \equiv (\hat{q}_1, \hat{p}_1, \ldots, \hat{q}_N, \hat{p}_N)^T$.

We consider Gaussian unitary operations of the form $U = {\rm exp}(-i\hat{H}/2)$ where $\hat{H}$ is \textit{bilinear} in the field operators. Then, in the Heisenberg picture, such operations transform $\vu{a} \to U^\dagger\vu{a} U$, or equivalently, transform the quadrature operators via $\vu{x} \to \vb{S}\vu{x}$. In order to respect the canonical commutation relations, this real $2N\times 2N$ matrix $\bm{S}$ must be symplectic: $\bm{S}\bm{\Omega}\bm{S}^T = \bm{\Omega}$, where $\bm{\Omega}$ is block diagonal: 
\begin{equation}
    \bm{\Omega} = \bigoplus_{i=1}^N \bm{\omega} = {\rm diag}(\bm{\omega}, \ldots, \bm{\omega}) \quad \text{with}\quad \bm{\omega} = \begin{pmatrix}
    0 & 1 \\ - 1 & 0
    \end{pmatrix}
\end{equation}
In the main text, we refer to single-mode transformations as `local'; these correspond to $2\times 2$ symplectic matrices. We also use the label `local' to mean the direct sum of $N$ single-mode operations. 

The local transformation corresponding to phase-space rotation (i.e. phase shifting) is given by $\hat{R}(\theta) = \exp[-i\theta \hat{a}^\dagger \hat{a}]$, represented in the quadrature basis by 
\begin{equation}
    \bm{R}(\theta) = \begin{pmatrix}
    \cos\theta & \sin\theta \\ -\sin\theta & \cos\theta
    \end{pmatrix}.
\end{equation}
We also make use of single-mode squeezing $\hat{Z}(r) = \exp[r(\hat{a}^2 - \hat{a}^{\dagger 2})/2]$, given in the quadrature basis by 
\begin{equation}
    \bm{Z}(r) = \begin{pmatrix}
    e^{-r} & 0 \\ 0 & e^r
    \end{pmatrix}.
\end{equation}

\section{Construction of local operations}

In this section, we will elaborate on the existence of local operations in our protocols and the various constraints these operations must satisfy. Particularly, we will see why we can implement our protocol without squeezing each of the involved modes. For concreteness and without loss of generality, let us consider the example of two-mode decoupling, starting from the generic interaction
\begin{equation}
\bm{S} = \begin{pmatrix}
S_{11} & S_{12} & S_{13} & S_{14}\\
S_{21} & S_{22}  & S_{23} & S_{24}\\
S_{31} & S_{32} & S_{33} & S_{34}\\
S_{41} & S_{42} & S_{43} & S_{44}\\
\end{pmatrix}.
\end{equation}
We also refer to the rows and columns of this matrix as $\bm{S} = (\vb{u}_1, \vb{v}_1, \vb{u}_2, \vb{v}_2)^T$ and $\bm{S} = (\vb{x}_1, \vb{y}_1, \vb{x}_2, \vb{y}_2)$, respectively. 

  The function of each local operation in the protocol is essentially to align two originally nonparallel 4-dimensional vectors (in the two-mode case). For instance, the first recursive step of the decoupling scheme involves choosing a local operation $\bm{L}^{(q)}$ such that $\bm{L}^{(q)}\vb{x}_1 =-\bm{\Omega} \vb{u}_1$. Note that each two-dimensional subvector can be transformed independently to the target subvector using a single-mode Gaussian operation, e.g. $ \bm{L}_1^{(q)}(S_{11}, S_{21})^T = \bm{\omega}(S_{11}, S_{12})^T = (S_{12}, -S_{11})^T$, which is constructed via
  \begin{align}
\begin{split}
   \bm{L}_1^{(q)} &= \bm{R}(-\theta)\bm{Z}(r)\bm{R}(\varphi) \\ &= \begin{pmatrix}
-\sin \theta & -\cos\theta \\ \cos \theta & -\sin\theta
\end{pmatrix}\begin{pmatrix}
e^{-r} & 0 \\ 0 & e^{r}
\end{pmatrix}\begin{pmatrix}
\cos\varphi & \sin\varphi \\ -\sin\varphi & \cos\varphi
\end{pmatrix},
\end{split}
\label{eq:2mode-decomp}
\end{align}
where $\theta = \arctan(-S_{11}/S_{12})$, $ r = \ln(\frac{\sqrt{S_{11}^2 + S_{21}^2}}{\sqrt{S_{11}^2 + S_{12}^2}}) $ and $\varphi = \arctan(S_{21}/S_{11})$. Similarly, we would transform $\bm{L}_2^{(q)}(S_{31}, S_{41})^T = \bm{\omega}(S_{13}, S_{14})^T$ using a similar combination of phase-shifting and local squeezing.  

Clearly the choice of local operations above is not unique. In fact, it turns out that we can relax the squeezing requirement on one of the two modes. To see this, we note that the first recursive step of our protocol only requires use of the orthogonality relations. That is, it suffices to transform $\bm{L}^{(q)}\vb{x}_1 =-c\,\bm{\Omega} \vb{u}_1$, up to a rescaling factor $c$, since this factor does not change the orthogonality between rows/columns. In the choice of $\bm{L}^{(q)}$ above, we have taken $c = 1$. However, we can also, for instance, relax the squeezing requirement on mode 1, i.e. set $r \to 1$ in Eq.~\eqref{eq:2mode-decomp} above. To compensate, we then correspondingly need additional local squeezing on mode 2, so that overall $c = \norm{(S_{11}, S_{12})}$. In general, any local operation must be chosen so that the ratios of the norms of the two-dimensional subvectors are preserved, i.e. so that 
\begin{equation}
    \frac{\norm{(-S_{12}, S_{11})^T}}{\norm{\bm{L}_1^{(q)}(S_{11}, S_{21})^T}} = \frac{\norm{(-S_{14}, S_{13})^T}}{\norm{\bm{L}_2^{(q)}(S_{31}, S_{41})^T}}
\end{equation}
For the case of $c = 1$, the left-hand side and right-hand side of this equality both have value 1. This is not a necessary condition, and so the two sides can take any (finite nonzero) value so long as they are equal. 

Thus, it is possible to redefine the local operations in the first recursive step up to a rescaling factor such that $\bm{L}_1^{(q)}$ is completely squeezing free (though $\bm{L}_2^{(q)}$ is generally not). As a result, for the second recursive step, we also need to choose $\bm{L}^{(p)}\bm{\gamma}_1 = -(c + c^{-1}) S'_{22}  \bm{\Omega} \bm{\alpha}_1 + c^2 \bm{\Omega} \bm{\beta}_1$, where $\bm{S'} = (\bm{\alpha}_1, \bm{\beta}_1, \bm{\alpha}_2, \bm{\beta}_2)^T = (\bm{\chi}_1, \bm{\gamma}_1, \bm{\chi}_2, \bm{\gamma}_2)$ as given in the main text. This new transformation can still be realized by modifying $\bm{L}^{(p)}_2$ while keeping $\bm{L}^{(p)}_1 =-\bm{\omega} = \bm{R}(-\pi/2)$. In this case, then, we see that both $\bm{L}_1^{(q)}$ and $\bm{L}_1^{(p)}$ require \textit{no} squeezing and so, in effect, we have eliminated the requirement of local squeezing for mode 1 entirely.

The same relaxation holds for the two-mode transduction example as well. Here, in the first recursive step, we choose an $\bm{M}^{(q)}$ such that $\bm{M}^{(q)}\vb{x}_1 =-\bm{\Omega} \vb{u}_2$, as is shown in the main text. However, this only needs to satisfy the orthogonality relations, and so holds up to a constant. Thus, we can generally make $\bm{M}_1^{(q)}$ squeezing free, and as shown in the text, we always pick $\bm{M}_1^{(p)} = -\bm{\omega}$. Thus, our two-mode transduction protocol likewise requires squeezing only one of the modes. 

In the multimode case, a similar argument holds. Here, for $N$ mode, we must transform one $2N$-dimensional vector to another. We can thus relax squeezing on one of the modes, and correspondingly modify the intermediate local operations for the remaining $N-1$ modes, so as to maintain the ratios of each of the subvectors. Thus, in general, for an $N$-mode interaction, our protocol requires local squeezing on only $N-1$ of the modes.

\section{Strategy for the edge cases}
To decouple or swap modes, we assumed that the symplectic matrix $\bm{S}$ is generic such that the required local operations always exist. However, there are situations where this is not possible --- specifically, when either the initial or final quadrature vectors is a zero vector. In this section, we explain a strategy for converting such an edge case into the generic case for which our protocols can be applied.

Before proceeding, we have to point out a subtle distinction between the decoupling protocol and the transduction protocol. Although the strategy introduced in this section can be applied to decoupling and swapping modes, the applicability of the transduction protocol requires some extra discussion. Firstly, it is easy to see that if a symplectic matrix is a permutation of modes, up to local operations, then whether it can be used to apply the transduction protocol entirely depends on the structure of the permutation. For the chosen pair of modes, one can implement the transduction protocol using the permutation-like symplectic matrix if and only if the corresponding permutation contains an order-2 cycle swapping these two modes. A quick counterexample is that a local operation, which can be viewed as a trivial permutation, obviously cannot be used to implement the transduction protocol. This observation may be further generalized to more complicated symplectic matrices, which will be discussed in more detail in our future follow-up works.

\textit{Two-mode situation--} We start with the two-mode decoupling case. An edge case happens when the local operation cannot be constructed using the geometric picture. For example, as shown in the main text, the local operation $\bm{L}_1^{(q)}$ exists only if  $(S_{11}, S_{21} )$ and $(S_{11}, S_{12})$ are either both zero (in which case apply the identity) or both nonzero (in which case apply $\bm{L}_1^{(q)}$ as prescribed). A non-generic interaction is where only one out of a pair of subvectors is nonzero, so that the denominator of the corresponding local operation is vanishing. Such an interaction must contain a two-by-two block
\begin{equation}
\begin{pmatrix}
S_{2k-1,2l-1} & S_{2k-1,2l} \\
S_{2k,2l-1} & S_{2k,2l}
\end{pmatrix}
\end{equation}
for some $k,l\in \{1,2\}$, with a certain zero row or column. Since we can always apply arbitrary single-mode operation before and after $\bm{S}$ to modify the matrix elements, we must then further assume that the above two-by-two block is actually a zero matrix. Evidently, in order to make it an edge case, we will require
\begin{equation}
\begin{pmatrix}
S_{2l-1,2k-1} & S_{2l-1,2k} \\
S_{2l,2k-1} & S_{2l,2k}
\end{pmatrix}
\end{equation}
to be non-zero. Then for the two-mode edge case, without loss of generality, we consider the interaction
\begin{equation}
\bm{S} = \begin{pmatrix}
S_{11} & S_{12} & 0 & 0\\
S_{21} & S_{22}  & 0 & 0\\
S_{31} & S_{32} & S_{33} & S_{34}\\
S_{41} & S_{42} & S_{43} & S_{44}\\
\end{pmatrix}.
\end{equation}
However, by the definition of symplectic matrix, $\bm{S} \bm{\Omega} \bm{S}^T = \bm{\Omega}$ (and $\bm{S}^T \bm{\Omega} \bm{S} = \bm{\Omega}$ equivalently), one can easily verify that the submatrix  $ \begin{psmallmatrix}
S_{31} & S_{32} \\ S_{41} & S_{42}
\end{psmallmatrix}$ must also be zero, meaning that $\bm{S}$ is already a decoupled. Therefore, the two-mode edge cases will not increase the number of copies of $\bm{S}$ required to decouple a bosonic mode.

\textit{Multimode situation--} Likewise for the multimode decoupling case, our strategy for constructing the local operations works when each single-mode projection of the rows and columns to be either both zero or both nonzero, because we cannot use local operations to transform a nonzero/zero vector to a zero/nonzero vector.  For the same reason explained in the two-mode case, without loss of generality, an $N$- mode edge-case interaction $\bm{S}$ should at least contain a zero submatrix, for example $\begin{psmallmatrix}
S_{1,3} & S_{1,4} \\ S_{2,3} & S_{2,4}
\end{psmallmatrix} $ (meanwhile with $\begin{psmallmatrix}
S_{3,1} & S_{4,1} \\ S_{3,2} & S_{4,2}
\end{psmallmatrix} \neq 0$). We will also randomize each non-zero two-by-two block of $\bm{S}$ by a pre-local operation $\bm{L}^{(R)}$ and a post-local operation $\bm{M^{(R)}}$ to make it full-ranked, i.e. $\bm{S}^{\rm new} = \bm{L}^{(R)}\bm{S}\bm{M}^{(R)} $, with  the local operations $\bm{L}^{(R)}, \bm{M}^{(R)}$ randomly chosen. Then it is easy to check that the number of the vanishing two-by-two blocks in the product $(\bm{L}^{(R)}\bm{S}\bm{M}^{(R)})^K$ will only decrease before the resulting net interaction is generic, if $\bm{S}$ is not a permutation of the bosonic modes. Since the number of the vanishing two-by-two blocks should not exceed $N(N-1)$ (because the a scattering matrix should always be full-ranked) and the $N$-th product of any permutation is the identity operation, we conclude there always exists an integer $K\leq N(N-1)$ such that the net interaction is generic even though the initial $\bm{S}$ is non-generic.

\section{Local operations for multimode transduction}

We first relabel the $N$ modes so that the original $N$-th mode is now the first mode, and the original first mode is now the $N$-th mode. Then the the local operation $\bm{L}'^{(q)}$ can be calculated using the same equation for $\bm{L}^{(q)}$ only with $\bm{S}$ replaced by $\bm{S}^{\rm dc}$. The local operation $\bm{M}'^{(p)}$ can be calculated similarly using the formula of $\bm{M}^{(p)}$ if we replace $\bm{S}^\ast$ with $\bm{S}^{\rm td} \bm{M'}^{(q)} \bm{S}^{\rm dc}$ and $\bm{S'}$ with $\bm{S}^{\rm dc} \bm{L'}^{(q)} \bm{S}^{\rm dc}$. The remaining local operation $\bm{M'}^{(q)}$ can be calculated by the following equations:

\begin{equation}
\big(\bm{M'}^{(q)}_k\big)_{ij} =
\frac{(-1)^{i+1} S_{3 \underline{i}} S_{\bar{j} 1}}{ (S_{\bar{i} 1})^2 + (S_{\underline{i}1})^2} -  \frac{(-1)^{j+1}  S_{3 \bar{i}} S_{\underline{j} 1}}{ (S_{3\bar{j}})^2 + (S_{3\underline{j}})^2}
\end{equation}
for $k<N$, and

\begin{equation}
    \bm{M'}^{(q)}_N = {\rm Id}.
\end{equation}

Note that these constructions are not unique and our previous discussions on the local operations and edge cases also apply in this situation.

\bibliography{bibliography}